\documentclass[journal]{IEEEtran}

\ifCLASSINFOpdf

\else

\fi

\usepackage{graphicx}

\usepackage{amssymb}
 \usepackage{amsthm}
\usepackage{amsmath}
\usepackage{color}
\usepackage{setspace}
\usepackage{algorithm}
\usepackage{algorithmic}
\usepackage{graphicx}
\usepackage{subfigure}
\usepackage{multirow}
\usepackage{array}
\usepackage[font=small,labelfont=bf,tableposition=top]{caption}
\usepackage{booktabs}
\usepackage{threeparttable}
\usepackage{stfloats}
\usepackage{color}
\renewcommand{\baselinestretch}{1.0}

\newenvironment{myitemize}[1][]{
\begin{list}{$\bullet$}
    {
     \setlength{\leftmargin}{2mm}     %
     \setlength{\parsep}{1mm}         %
     \setlength{\topsep}{0mm}         %
     \setlength{\itemsep}{0mm}        %
     \setlength{\labelsep}{1.5mm}     %
     \setlength{\itemindent}{0mm}    %
     \setlength{\listparindent}{5mm} %
    }}
{\end{list}}

\begin{document}


\title{Hierarchy-Dependent Cross-Platform Multi-View Feature Learning for Venue Category Prediction}

\author{Shuqiang Jiang, \textit{Senior Member, IEEE}, Weiqing Min, \textit{Member, IEEE}, and Shuhuan Mei
\thanks{S. Jiang is with the Key Laboratory of Intelligent Information Processing, Institute of Computing Technology, Chinese Academy of Sciences, Beijing, 100190, China, and also with University of Chinese Academy of Sciences, Beijing, 100049, China email: sqjiang@ict.ac.cn.
W. Min is with the Key Laboratory of Intelligent Information Processing, Institute of Computing Technology, Chinese Academy of Sciences, Beijing, 100190, China, and also with State key Laboratory of Robotics, Shenyang Institute of Automation, Chinese Academy of Sciences, Shenyang, 110016, China. email:minweiqing@ict.ac.cn.
S. Mei is with Shandong University of Science and Technology, Shandong, 266590, China, and also an intern with the Key Laboratory of Intelligent Information Processing, Institute of Computing Technology, Chinese Academy of Sciences, Beijing, 100190, China shuhuan.mei@vipl.ict.ac.cn.}
}
\markboth{IEEE Transactions on Multimedia,~Vol.~X,
No.~XX,~Month~Year}{}



\maketitle

\begin{abstract}
In this work, we focus on  visual venue category prediction, which can facilitate various applications for  location-based service and personalization. Considering that the complementarity of different  media platforms, it is reasonable to leverage  venue-relevant  media data from different platforms to boost the prediction performance. Intuitively, recognizing one venue category involves multiple semantic cues, especially objects and scenes, and thus they  should contribute together to venue category prediction. In addition, these venues can be organized in a natural hierarchical structure, which provides  prior knowledge to guide venue category estimation. Taking these  aspects into account, we propose a  Hierarchy-dependent Cross-platform Multi-view  Feature Learning (HCM-FL) framework for venue category prediction from  videos by leveraging images from other platforms. HCM-FL  includes two major components, namely Cross-Platform Transfer Deep Learning (CPTDL) and Multi-View Feature Learning  with the  Hierarchical Venue Structure (MVFL-HVS). CPTDL is capable of reinforcing the learned deep network  from  videos using images from other platforms. Specifically, CPTDL first trained a deep network using  videos. These  images from other platforms are filtered by the learnt  network and these selected images are then  fed into this learnt  network to enhance it. Two kinds of  pre-trained networks  on the ImageNet and Places dataset are employed. Therefore, we can harness both object-oriented and scene-oriented deep features through these enhanced deep networks. MVFL-HVS is then developed to enable multi-view feature fusion. It is capable of embedding the hierarchical structure ontology to support  more discriminative joint feature  learning. We conduct the experiment on  videos from Vine and images from Foursqure. These experimental results demonstrate the advantage of our proposed framework in jointly utilizing multi-platform data, multi-view deep features and  hierarchical  venue structure knowledge.
\end{abstract}


\IEEEpeerreviewmaketitle

\section{Introduction}
Recently, visual geo-localization has received a significant amount of attention  in both computer vision and multimedia community \cite{Vo-RIM2GPS-ICCV2017,Weyand-PlaNet-arXiv2016,Nie-EMVU-MM2017,Zhang-VCEMV-MM2016,Trevisiol-RGLV-ICMR2013,Choi-MMLE-Springer2015} because of its  various applications, such as location based recommendation service \cite{Zah-IML-TMM2015}, augmented reality \cite{He-AR-CNN2016} and photo forensics\footnote{https://traffickcam.com/about}. One task of visual geo-localization is visual venue category prediction and its  goal is to predict the venue category (e.g., Dessert Shop and Pet Store) from images or videos. It is especially important in  social media applications  such as venue recommendation \cite{Zah-IML-TMM2015} and  tourist route planning. For example  we could  recommend him/her a particular venue (e.g.,  Chinese restaurant and movie theater) and  give more accurate check-in suggestions based on predicted venue categories a user has visited. Therefore, in this work, we focus on visual venue category prediction.

There are some previous works on venue category prediction. For example, Chen \textit{et al.} \cite{Chen-MBVR-AAAI2016} mined business-aware visual concepts from social media to recognize the business venue  from  images. Zhang \textit{et al.} \cite{Zhang-VCEMV-MM2016} proposed a  tree-guided multi-task multi-modal learning approach  to jointly fuse multi-modal information from videos for venue category prediction. Recently, Nie \textit{et al.} \cite{Nie-EMVU-MM2017}  used  external sound knowledge to enhance the acoustic modality for  venue category estimation from videos. These works mainly utilized  information from a single platform for this task. However, little work has investigated this problem via exploiting  media data from other platforms, which is especially vital in  the deep learning era.
\begin{figure}
\centering
 \includegraphics[width=0.40\textwidth]{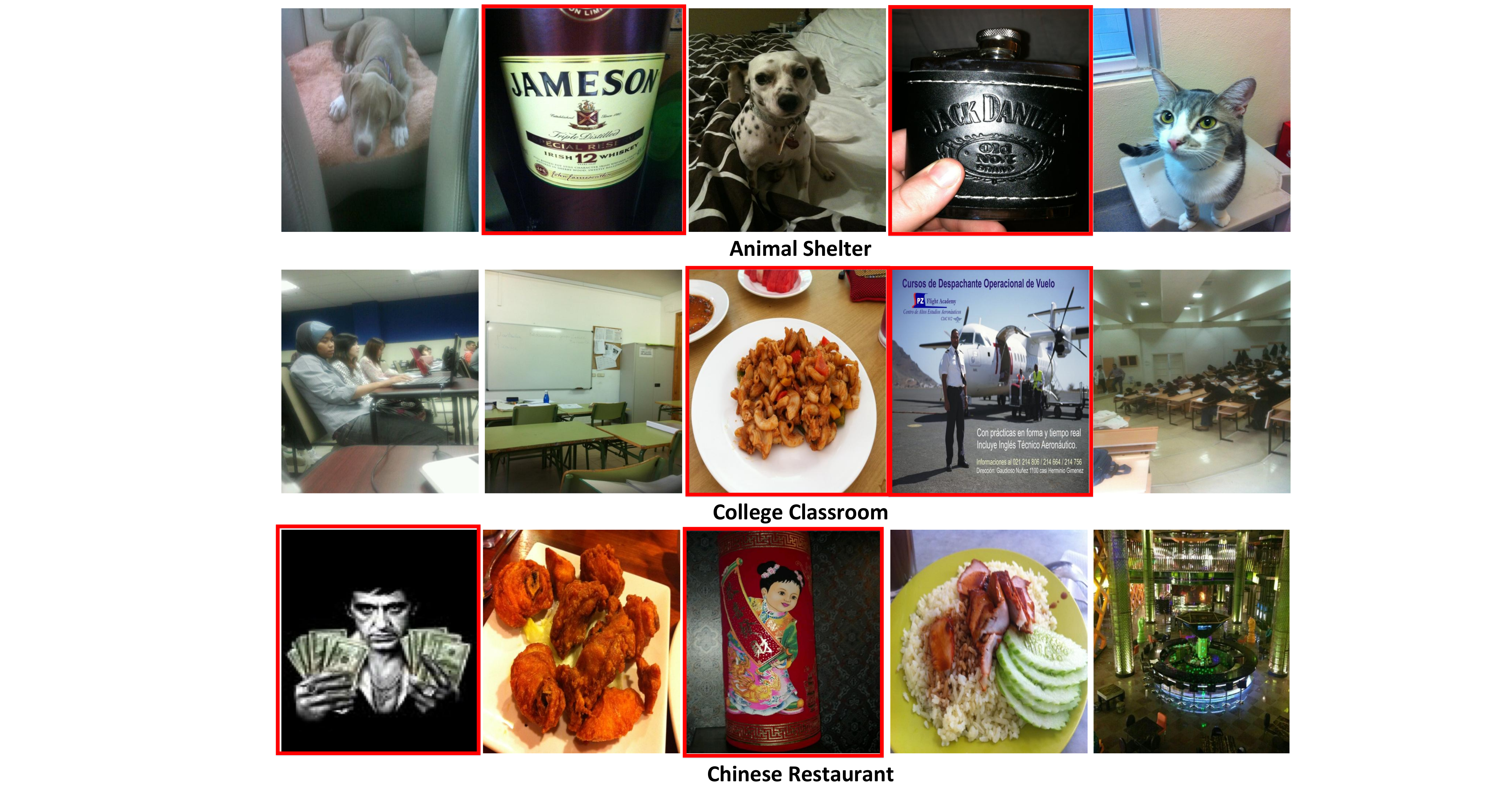}
\caption{Some venue categories  in Foursquare, where the images labeled with red boxes are noisy ones.}
\label{noise-hvs}
\end{figure}

With the success of photo-sharing social websites, we can easily crawl sufficient venue-annotated images  from various platforms, such as  Foursquare and Instagram. Meanwhile, motivated by the promising results of deep networks on visual analysis tasks, there have also been a number of attempts to utilize  deep networks for venue category recognition from videos \cite{Nie-EMVU-MM2017,Zhang-VCEMV-MM2016}. However, training such  deep  networks generally need  large-scale  data. Therefore, automatically sampling  more image data from other platforms  appears as a natural way to improve the prediction performance. For example, some works  \cite{Gan-YLWE-CVPR2016,Gan-WSVR-ECCV2016} have leveraged  web images and videos from  Google and Youtube for video classification. However, directly using  images from other platforms  probably hurts the performance for the following two reasons. First, these  images  from social websites are  noisy. Fig. \ref{noise-hvs} shows some examples of venue categories from Foursquare. We can see that there are many noisy ones, and the content of some images is irrelevant to the labeled venues.  In addition, there is a well-known  domain gap problem \cite{Saenko-AVC-ECCV2010} for the data from different platforms.

\begin{figure*}
\centering
 \includegraphics[width=0.75\textwidth]{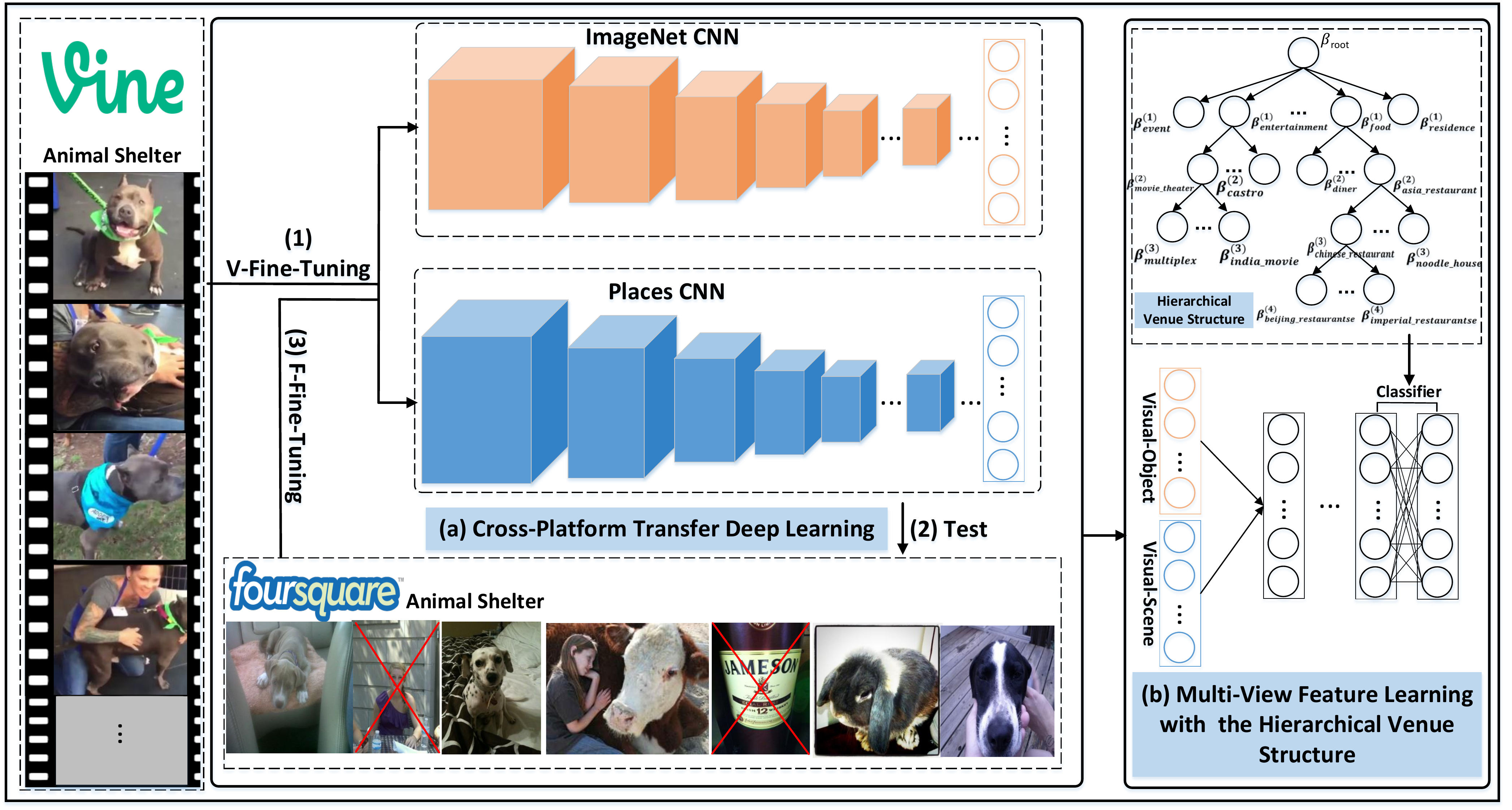}
\caption{The proposed Hierarchy-dependent Cross-platform Multi-view  Feature Learning (HCM-FL) framework.}
\label{HCM_FL Framework}
\end{figure*}
\begin{figure}
\centering
 \includegraphics[width=0.35\textwidth]{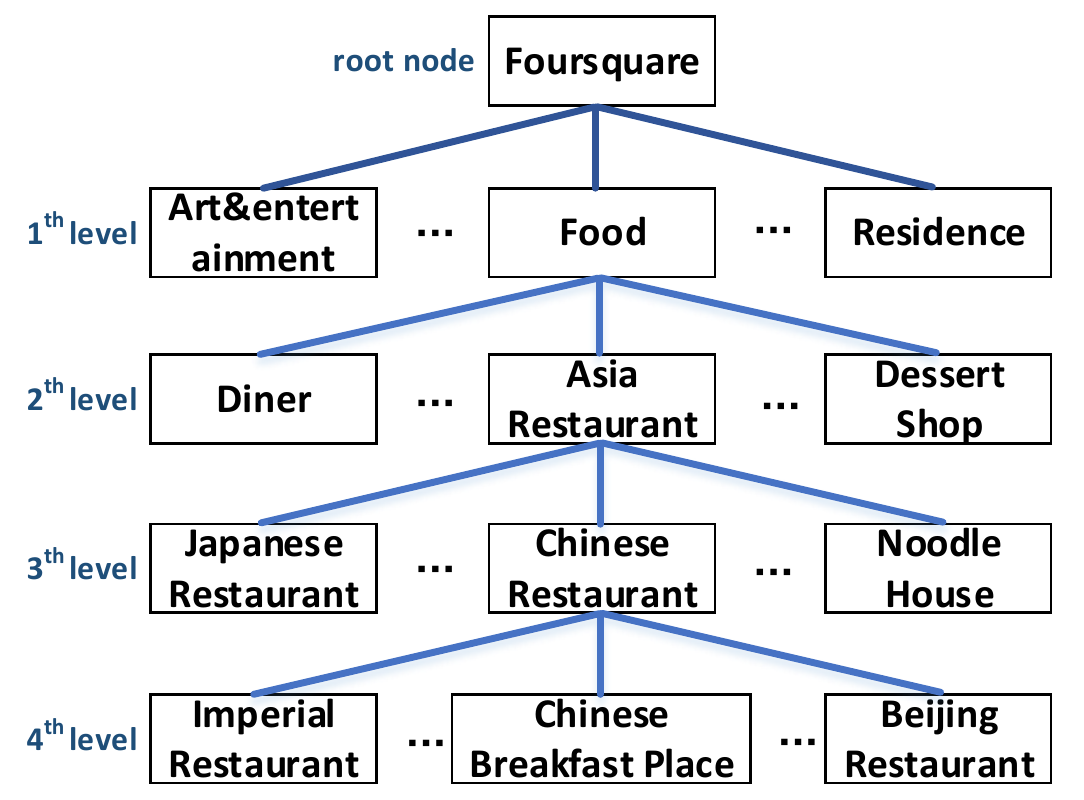}
\caption{The hierarchical structure of the venue categories in Foursquare. We  illustrate a part of the structure
due to the limited space.}
\label{example-hvs}
\end{figure}

Intuitively, recognizing a venue category involves  various semantic cues, especially objects and scenes. For example, if there is a dog in one video, then some venue categories such as  "Animal Shelter" and "Pet Store" become a probable one. If one video  is about the  "indoor scene", then the probability of some venue categories such as  "Monument" and "Road" reduces. Therefore, both  scene and object semantics provide  strong  context for venue category prediction. Another point to consider is that there are sometimes natural hierarchical structures for venues. For example, the venues in Vine\footnote{The associated venues of the videos in vine are mapped to  venue categories in Foursquare.} are organized in a four-layer tree structure\footnote{https://developer.foursquare.com/docs/resources/categories}. Fig. \ref{example-hvs} shows a small part of this structure. Knowing the venue structure allows us to borrow the  knowledge from relevant venue categories to learn more discriminative features, especially for  venue categories with less samples. For example, it is likely that  these uploaded videos in the Japanese Restaurant are more similar to ones  in  the Seafood Restaurant than those in the Christmas Market. In addition, the  prediction performance is affected by  unbalanced samples on different venue categories. For example,  there might be many videos for the theme park  but less ones for the Christmas Market in Vine.

Taking all the above-mentioned factors into consideration, we propose a  Hierarchy-dependent Cross-platform Multi-view  Feature Learning (HCM-FL) framework  for venue category prediction from  videos. As shown in Fig. \ref{HCM_FL Framework}, HCM-FL  mainly consists of two components: (a) Cross-Platform Transfer Deep Learning (CPTDL) (b)  Multi-View Feature Learning with the  Hierarchical Venue Structure (MVFL-HVS). In particular, we take  two platforms Vine and Foursquare in our study and focus on predicting venue categories from videos in Vine by exploiting  images from Foursquare. CPTDL first trained a deep network using  videos in Vine. The  images from Foursquare are filtered by the learnt network and these selected images are then  fed into this learnt network to enhance it. Two kinds of pre-trained networks are employed based on ImageNet1000 \cite{Alex-ImageNet-NIPS2012} and Places205 \cite{Zhou-LDFSR-NIPS2014}. Therefore, we can harness  both object-oriented and scene-oriented deep features through these two kinds of deep networks enhanced by CPTDL, respectively. MVFL-HVS is then developed to learn  joint  representation from these two types of features. Furthermore, MVFL-HVS can embed the hierarchical structural ontology among venue categories  to make learned  joint features more discriminative.

The contributions of our paper can be summarized as follows:
\begin{myitemize}
\item To our knowledge, this is the first study of cross-platform based  venue category prediction from videos, where we utilized the transfer deep learning method to  enhance the trained network from videos in one platform  by  taking full advantage of  images from the other platform.
\item We proposed a hierarchy-dependent cross-platform multi-view  feature learning  framework for venue category prediction. In this framework, we further developed a  multi-view feature learning network, which can  embed  the hierarchical venue structure knowledge to enable more discriminative joint feature learning.
\item We conducted the experiment on  two platforms Vine and Foursquare, and these experimental results  validated the effectiveness of our proposed framework in fully utilizing  multi-platform data, object-scene semantic features and  hierarchical  venue structure knowledge.
\end{myitemize}

\section{The Proposed Framework}

As shown in Fig. \ref{HCM_FL Framework}, in this section, we introduce our proposed  Hierarchy-dependent Cross-platform Multi-view  Feature Learning (HCM-FL) framework, which mainly consists of two components, namely (a) Cross-Platform Transfer Deep Learning (CPTDL) and (b)  Multi-View Feature Learning with the  Hierarchical Venue Structure (MVFL-HVS).  As mentioned before, our task is to predict venue categories from videos in Vine by exploiting images from Foursquare. CPTDL  first   uses  images and videos from two platforms to reinforce the training on two kinds of pre-trained deep networks, namely ImageNet CNN and Places CNN, respectively. Based on two kinds of enhanced networks, we harness both object-oriented and scene-oriented deep visual features for each video. MVFL-HVS is then developed to fuse these two kinds of deep features into a unified feature representation. By embedding the hierarchical venue structure, the fused features are more discriminative. We next introduce each component in details.

\subsection{Cross-Platform Transfer Deep Learning (CPTDL)}
As shown in Fig. \ref{HCM_FL Framework}(a), we adopt the   VGG-16 deep network  \cite{Simonyan-VDCN-arXiv2014} as the basic architecture in CPTDL. Particularly, we use two kinds of pre-trained  VGG-16 networks: ImageNet CNN, pre-trained on the ImageNet1000 dataset \cite{Alex-ImageNet-NIPS2012}, and Places CNN, pre-trained on the  Places205 dataset \cite{Zhou-LDFSR-NIPS2014}. ImageNet CNN is mainly used to extract the visual object features while Places CNN is mainly used to extract  visual scene features \cite{Herranz_SRCNN_CVPR2016}.  They are complementary and contribute together  to venue category prediction.

CPTDL can enhance ImageNet CNN and Places CNN, respectively. We take the ImageNet CNN as an example to describe the training process of CPTDL. Since our task is to predict venue categories from videos in Vine, we start by training a network  using the training set of videos. For this training, each video is decomposed into a set of key frames. We first use these key frames to  fine-tune  the ImageNet CNN, that is V-Fine-Tuning.  Then this fine-tuned network is used to test images from Foursquare to filter out noisy images. By utilizing the remaining images from Foursquare to further fine-tune this network, namely F-Fine-Tuning, we can obtain the final enhanced ImageNet CNN.

Note that after V-Fine-Tuning,  we next should utilize the images from Foursquare to improve the fine-tuned ImageNet CNN. Although  related Foursquare images  are helpful for venue prediction, there are usually noisy ones in Foursquare. Take the class ``Animal Shelter" as an example (Fig. \ref{HCM_FL Framework}(a)), some images do not describe this venue category. In order to remove useless Foursquare images and keep related ones, we use the ImageNet CNN after V-Fine-Tuning to perform filtering. Formally, for $(x_{m}, t_{m})$, where $x_{m}$ denotes the $m$-th image from Foursquare and $t_{m}\in T$ is its category label. $|T|$ is the number of venue categories. Each image $x_{m}$ is fed into the fine-tuned ImageNet CNN in a feed-forward way, and yields a probability distribution $p_{m}\in {R}^{|T|}$ over the $|T|$ video venue categories. We use $p_{m}(t)$ to denote the probability of image $m$ belonging to  the $t$-th category. We keep the image $m$ as the $t_{m}$-th category if  $p_{m}(t_{m})$ is in the top $K$-ranked probability, where $K$ is a threshold, and $K=100$ in our experiment. The cleaned Foursquare images are then used to further fine-tune the ImageNet CNN  and obtain the final network, which  focuses more on video venue categories enhanced by cleaned Foursquare images.

We can adopt a similar training process to obtain the enhanced Places CNN.  We then extract  visual object features and visual scene features  for key frames from training videos based on the  enhanced ImageNet CNN and Places CNN, respectively.

\subsection{Multi-View Feature Learning with the  Hierarchical Venue Structure(MVFL-HVS)}
Through CPTDL, for each key frame from each training video, we obtain visual object  features and scene features   from enhanced ImageNet CNN and Places CNN, respectively. In order to fuse these two kinds of features, we first  transform visual features from key frames of videos to the video-level feature representation. Particularly, for visual object features, we  use  the output of the  FC7 layer from the enhanced ImageNet CNN  as the input to the fusion network. That is, for the $j$-th key frame of video $i$, $\mathbf{f}_{i,j}$, this pathway outputs $\mathbf{f}_{i,j} \mapsto \mathbf{x}^{O}_{i,j}\in \mathbb{R}^{4096}$. After mean pooling, we obtain the feature representation $\bar{\mathbf{x}}^{O}_{i} = \sum_{j=1}^{n_{i}}x^{O}_{i,j}$, where $n_{i}$ is the number of key frames for video $i$. We adopt a similar strategy and  use  the output of the  FC7 layer from the enhanced Places CNN  to obtain visual scene features  $\bar{\mathbf{x}}^{S}_{i} = \sum_{j=1}^{n_{i}}x^{S}_{i,j}$ for each training video. These two kinds of features are further used as the input to learn a  multi-view  feature fusion network via MVFL-HVS (Fig. \ref{HCM_FL Framework}(b)).

Different from  existing supervised multi-view deep feature fusion networks, MVFL-HVS can embed the hierarchical venue structure ontology to support more discriminative joint feature learning.  The venues of videos in Vine are organized in a  hierarchical way with the four-layer ontology (Fig. \ref{example-hvs}).  These Hierarchical Venue Structures (HVS) can be used to guide the venue category prediction. As shown in Fig.\ref{HCM_FL Framework}(b), we  next introduce how  to utilize the hierarchical structure prior into our multi-view feature learning network via MVFL-HVS.

We define a hierarchy as a set of nodes $\Omega=\{1,2,...\}$ with the parent relationship $\pi : \Omega \rightarrow \Omega$, where $\pi(n)$ is the parent of node $n \in \Omega$. $C_{n}$ is the set of all the children of node $n$. ${D}=\{(f_{\mathbf{w}}(\mathbf{x}_{m}),t_{m})\}_{m=1}^{M}$ denotes the training data, where $x_{m}$ is an instance, $f_{\mathbf{w}}(\mathbf{x}_{m}) \in {\mathbb{R}}^{d}$ is the transformed representation through our multi-view fusion network with  parameters $\mathbf{w}$. In our work, a multi-layer feedforward neural network is adopted as the fusion network. $t_{m}\in T$ is its venue category. $d$ is the dimension of transformed features. $T \in \Omega$ is the set of leaf nodes in the hierarchy labeled from 1 to $|T|$. $M$ is the size of training set. We assume that each instance is assigned to one of leaf nodes in the hierarchy. The top-level weight parameters of the last fully-connected classifier layer from the fusion network is$\{\pmb \beta_{n}\}^{|T|}_{n=1}$, where $\pmb \beta_{n} \in \mathbb{R}^{d \times 1}$.
\begin{figure}
\centering
\includegraphics[width=0.30\textwidth]{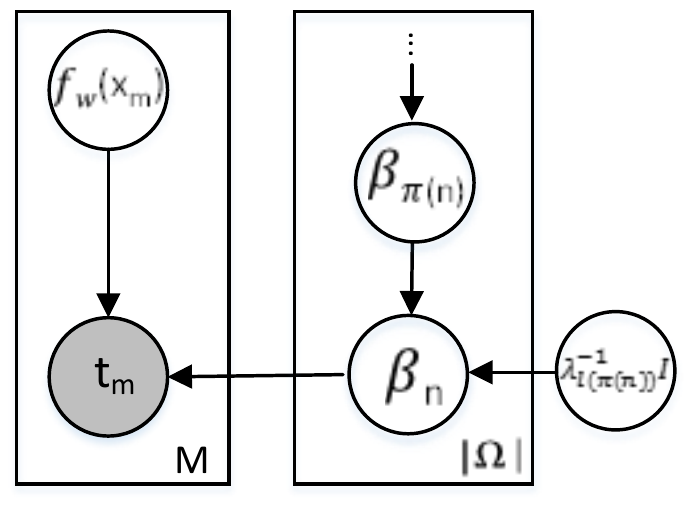}
\caption{The graphical representation of the hierarchical structure model.}
\label{graph_lhvs}
\end{figure}

 The parent-child relationship is modeled by placing a hierarchical prior over the children nodes centered around the parameters of their parents. Therefore, it can encourage  venue categories nearby in the hierarchy to share similar model parameters. The hierarchical  graph representation is shown in Fig. \ref{graph_lhvs} and  the joint probability distribution is
\begin{equation}\label{beta_final_loss}
\begin{aligned}
&p(\mathbf{D},\pmb \beta,\mathbf{w})=\\
&\underset{m}{\prod}p(t_{m}|f_{\mathbf{w}}(\mathbf{x}_{m}),\{\pmb \beta_{n}\}_{n \in T})\underset{n \in \Omega}{\prod}p(\pmb \beta_{n}|\pmb \beta_{\pi(n)},\lambda^{-1}_{l(\pi(n))}\mathbf{I})\\
\end{aligned}
\end{equation}
where $\pmb \beta = \{\pmb \beta_{n}\}_{n \in \Omega}$. $\lambda^{-1}_{l(\pi(n))}I$ is  a diagonal covariance matrix with the diagonal element $\lambda^{-1}_{l(\pi(n))}$. $\lambda_{l(\pi(n))}$  are hyper-parameters. $l(\pi(n))$ is a mapping function, which maps the index of the node to the corresponding layer/level.
\begin{equation}\label{beta_final_loss}
\begin{aligned}
&p(\pmb \beta_{n}|\pmb \beta_{\pi(n)},\lambda^{-1}_{l(\pi(n))}\mathbf{I})=\mathcal{N}(\pmb \beta_{n}|\pmb \beta_{\pi(n)},\lambda^{-1}_{l(\pi(n))}\mathbf{I})\\
\end{aligned}
\end{equation}
where $\mathcal{N}(\cdot)$ is the Gaussian distribution with the mean $\pmb \beta_{\pi(n)}$ and the covariance matrix $\lambda^{-1}_{l(\pi(n))}\mathbf{I}$.

The posterior distribution is $p(\pmb \beta,\mathbf{w}|\mathbf{D}) \propto p(\mathbf{D},\pmb \beta,\mathbf{w})$. We then use  the Maximum A Posteriori probability (MAP) to estimate parameters $\{\mathbf{w},\pmb \beta\}$.

\begin{equation}\label{final_loss}
\begin{aligned}
&\log p(\pmb \beta,\mathbf{w}|\mathbf{D}) \propto \sum_{m}\log p(t_{m}|f_{\mathbf{w}}(\mathbf{x}_{m}),\{\pmb \beta_{n}\}_{n \in T})\\
&+\sum_{n \in \Omega}[-\frac{d}{2}\log 2\pi +\frac{d}{2}\log \lambda_{l(\pi(n))}]\\
&-\sum_{n \in \Omega}[\frac{\lambda_{l(\pi(n))}}{2}(\pmb \beta_{n}-\pmb \beta_{\pi(n)})^\top (\pmb \beta_{n}-\pmb \beta_{\pi(n)})]\\
\end{aligned}
\end{equation}
That is, we should minimize
\begin{equation}\label{final_loss}
\begin{aligned}
&\mathcal{L}(\pmb \beta,\mathbf{w})= -\sum_{m}\log p(t_{m}|f_{\mathbf{w}}(\mathbf{x}_{m}),\{\pmb \beta_{n}\}_{n \in T})\\
&+\sum_{n \in \Omega}[\frac{\lambda_{l(\pi(n))}}{2}(\pmb \beta_{n}-\pmb \beta_{\pi(n)})^\top (\pmb \beta_{n}-\pmb \beta_{\pi(n)})]\\
\end{aligned}
\end{equation}
where  the softmax classifier is used.

The loss function $\mathcal{L}(\pmb \beta,\mathbf{w})$ in Eq. \ref{final_loss} can be optimized by iteratively performing the following two
steps.

(i) Minimizing over $\{\pmb \beta_{n}\}_{n \in T}$ and $\mathbf{w}$  keeping $\{\pmb \beta_{n}\}_{n \in \Omega \texttt{\char92}T}$ fixed. This can be implemented using standard Stochastic Gradient Descent (SGD) algorithm to minimize
\begin{equation}\label{final_loss_E}
\begin{aligned}
&\min_{\mathbf{w},\{\pmb \beta_{n}\}_{n \in T}}\{-\sum_{m}p(t_{m}|f_{\mathbf{w}}(\mathbf{x}_{m}),\{\pmb \beta_{n}\}_{n \in T})\\
&\quad \quad \quad +\sum_{n \in T}\frac{\lambda_{l(\pi(n))}}{2}(\pmb \beta_{n}-\pmb \beta_{\pi(n)})^\top (\pmb \beta_{n}-\pmb \beta_{\pi(n)})\}\\
\end{aligned}
\end{equation}
(ii) Minimizing over $\{\pmb \beta_{n}\}_{n \in \Omega \texttt{\char92}T}$ keeping  $\{\pmb \beta_{n}\}_{n \in T}$ and $\mathbf{w}$ fixed. In this step,  we should minimize
\begin{equation}\label{final_loss_M}
\begin{aligned}
&\min_{\{\pmb \beta_{n}\}_{n \in \Omega \texttt{\char92}T}} \sum_{n \in \Omega \texttt{\char92}T}\frac{\lambda_{l(\pi(n))}}{2}(\pmb \beta_{n}-\pmb \beta_{\pi(n)})^\top (\pmb \beta_{n}-\pmb \beta_{\pi(n)})\\
\end{aligned}
\end{equation}

When $n \in \Omega \texttt{\char92}T$,
\begin{equation}\label{final_loss_M}
\begin{aligned}
&\frac{\partial \mathcal{L}(\pmb \beta,\mathbf{w})}{\partial \pmb \beta_{n} }= -\sum_{c \in C_{n}}\lambda_{l(n)}(\pmb \beta_{c}-\pmb \beta_{n})+\lambda_{l(\pi(n))}(\pmb \beta_{n}-\pmb \beta_{\pi(n)})\\
\end{aligned}
\end{equation}
Let $\frac{\partial \mathcal{L}(\pmb \beta,\mathbf{w})}{\partial \pmb \beta_{n} }=0$, we obtain
\begin{equation}\label{beta_final_loss_1}
\begin{aligned}
\pmb \beta_{n} =\frac{\sum_{c \in C_{n}}\lambda_{l(n)}\pmb \beta_{c}+\lambda_{l(\pi(n))}\pmb \beta_{\pi(n)}}{\sum_{c \in C_{n}}\lambda_{l(n)}+\lambda_{l(\pi(n))}}
\end{aligned}
\end{equation}

During a training epoch, the forward pass will generate the input $f_{\mathbf{w}}(\mathbf{x}_{m})$ for our own loss layer. We then optimize $\{\pmb \beta_{n}\}_{n \in \Omega}$ and $\mathbf{w}$ iteratively according to (i) and (ii). They are then taken back in the backward pass alongside the gradients with respect to its input.

Once the MVFL-HVS  is trained, we can predict the venue category of videos in the test stage.  After obtaining key frames from one test video, we first   extract object features and  scene features based on the ImageNet CNN and Places CNN enhanced by CPTDL. We then obtain each kind of  video features via mean-pooling on features from key frames. These two kinds of video features are then fed into the trained multi-view feature fusion network from MVFL-HVS  to  obtain its probability  distribution on all the  venue categories. The venue with the highest probability is selected as the venue category of this test video.

\section{Experiment}
\subsection{Dataset}
\begin{figure*}
\centering
\includegraphics[width=1.0\textwidth]{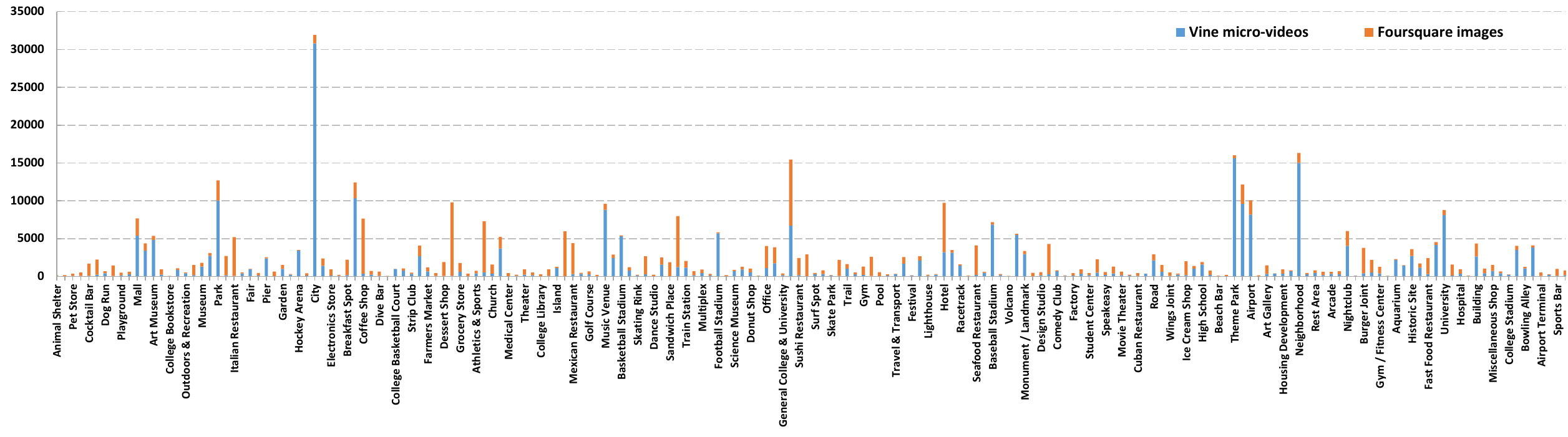}
\caption{The statistics of samples for 188 venue categories in two  platforms (best viewed under magnification).}
\label{sample-leaf-venue-sta}
\end{figure*}
\begin{figure}
\centering
 \includegraphics[width=0.40\textwidth]{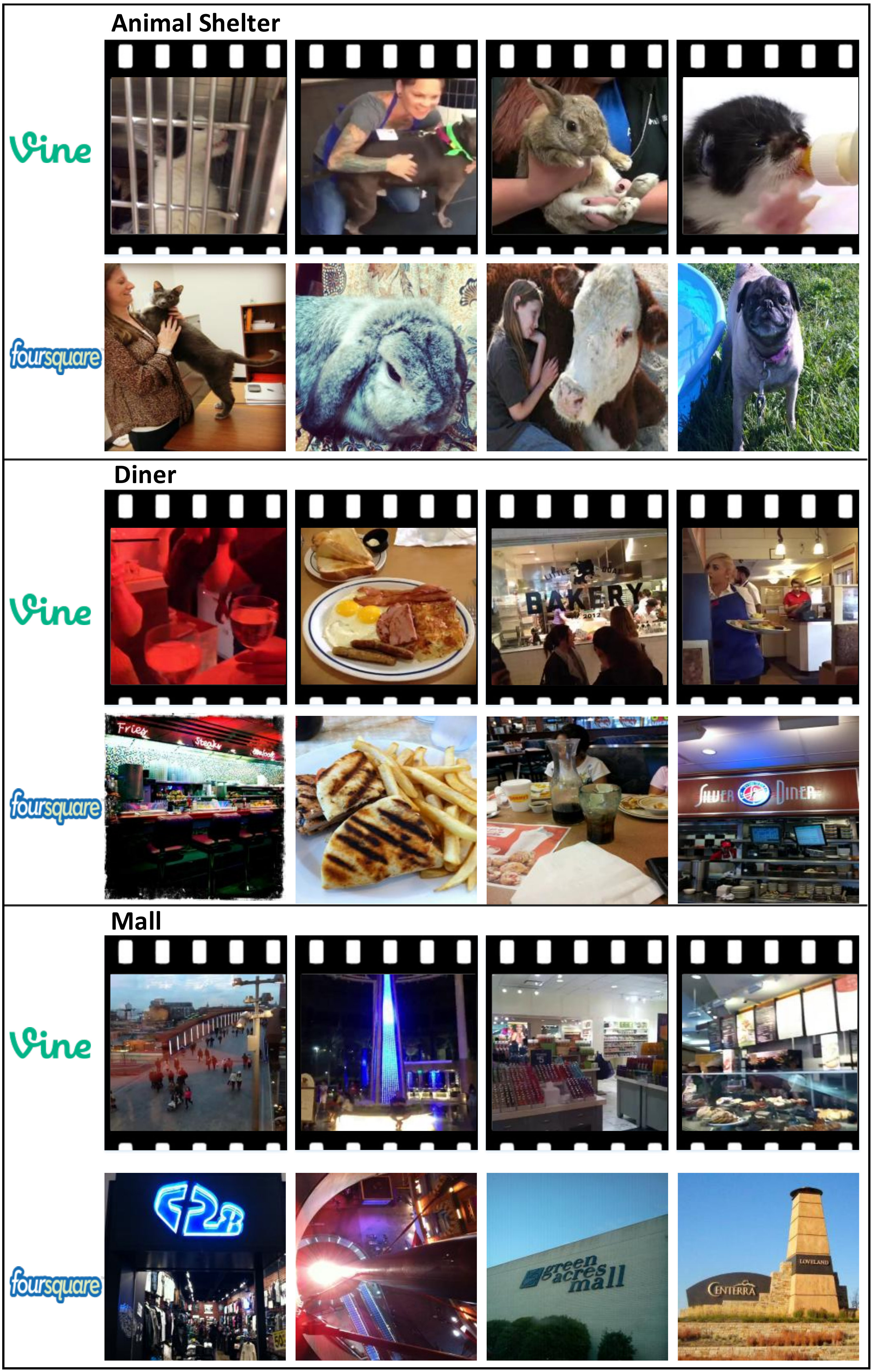}
\caption{Some example samples from Vine and Foursquare. Each row has 4 videos from Vine and 4 images from Foursquare with the same  venue labels.}
\label{exam-experiment}
\end{figure}
\textbf{Vine Dataset.} We use the dataset with  270,145 micro-videos from \cite{Zhang-VCEMV-MM2016}. Each
video is about 6 seconds in length. Many videos are automatically aligned with a venue category from Foursquare. There are totally 188 venue categories in this dataset. Foursquare organizes its venue categories into a four-layer hierarchical structure with 10, 389, 314 and 52  nodes in the first-layer, second-layer, third-layer and fourth-layer, respectively. Fig. \ref{example-hvs} shows a part of the hierarchical structure. Most of  venue categories in this dataset are in  the third layer, and the remaining ones are in the second layer.

\textbf{Foursquare Dataset.}  We use the Foursquare images from \cite{Min-CPMMTM-TMM2015}, where  each image is associated with  one venue label. We download images using the provided  urls and discard records for which venue labels do not belong to these 188 venue categories. The resulting Foursquare dataset consists of 190,299 images.

 Fig. \ref{sample-leaf-venue-sta}  provides statistics on both  videos from Vine and images from Foursquare while Fig. \ref{exam-experiment} shows some   videos and  images from three venue categories.

\subsection{Implementation Details}

In order to extract features from videos,  similar to \cite{Zhang-VCEMV-MM2016}, we  first select key frames from each video based on the color histogram. For each frame, we  calculate the L1 distance $l$ between  previous color histogram and current one. For all the calculated L1 distances of each video, we calculate their mean $\mu_{l}$ and variance $\sigma^{2}_{l}$. If $l>\mu_{l}+3\sigma_{l}$, this frame is marked as one candidate for key frames. If the number of candidates for one video is larger than 20, we select top $10$  candidates with larger differences as the key frames of this video; otherwise all the candidates are considered as the key frames.  The average of extracted  key frames for each video is 4  using this algorithm.

For the four-layer hierarchical structure in Foursquare, the annotated venue categories are in the third layer or the second layer. Therefore, our leaf node starts from the third layer. In addition, if there is one venue category of any videos assigned to the second layer, spawn a leaf-node under it and re-assign all the videos from this node of the second layer to this new leaf node \cite{Gopal-thesis2014}. Since we use  top-3 layers with additional root nodes (Fig. \ref{example-hvs}), we  define the distribution of parameters in the following forms:
\begin{equation}\label{beta_final_loss}
\pmb \beta_{n,l=3} \sim \mathcal{N}(\pmb \beta_{n,l=3}|\pmb \beta_{n,l=2},\lambda^{-1}_{2}\mathbf{I})
\end{equation}
\begin{equation}\label{beta_final_loss}
\pmb \beta_{n,l=2} \sim \mathcal{N}(\pmb \beta_{n,l=2}|\pmb \beta_{n,l=1},\lambda^{-1}_{1}\mathbf{I})
\end{equation}
\begin{equation}\label{beta_final_loss}
\pmb \beta_{n,l=1} \sim \mathcal{N}(\pmb \beta_{n,l=1}|\pmb \beta_{n,l=0},\lambda^{-1}_{0}\mathbf{I})
\end{equation}
where $\pmb \beta_{n,l=3}$ denotes the parameters from the nodes of the third layer (i.e., venue categories of videos) to the final layer of the fusion network. Generally, we set  $\pmb \beta_{n,l=0}=0$ and $\lambda_{0}=1$ because of their minimal effect on the remaining parameters \cite{Gopal-LSHC-NIPS2012,Gopal-thesis2014}. For other  hyper-parameters, considering the degree of the effect from different hierarchical layers, we empirically set $\lambda_{1}=5$, $\lambda_{2}=10$ in our experiment.


Similar  to \cite{Zhang-VCEMV-MM2016},  we randomly split our dataset into three subsets:  80\% of  videos  are used for training,  10\%  are used for validation, and the rest 10\% are used for testing. All the models are implemented on the  Caffe \cite{Jia-Caffe-arXiv2014} platform.

\subsection{Evaluation Metrics}
Similar to  \cite{Zhang-VCEMV-MM2016}, we use the standard  Macro-F1 and Micro-F1 metrics. Macro-F1 gives equal weight to each class-label in the averaging process; whereas  Micro-F1 gives equal weight to all instances in the averaging process. Both Macro-F1 and Micro-F1 reach their best score at 1 and worst one at 0. Let $TP_{t}$, $FP_{t}$, $FN_{t}$ denote the  true-positives, false-positives and false-negatives for the class-label $t \in T$.

The Macro-F1 and Micro-F1  are defined as follows \cite{Gopal-RRLSC-KDD2013}:
\begin{equation}
\begin{aligned}
&P_{t} = \frac{TP_{t}}{TP_{t}+FP_{t}},R_{t} = \frac{TP_{t}}{TP_{t}+FN_{t}}\\
&{\rm Macro-F1} = \frac{1}{|T|}\sum_{t}\frac{2P_{t}R_{t}}{P_{t}+R_{t}}\\
\end{aligned}
\end{equation}
\begin{equation}
\begin{aligned}
&\quad \quad \quad \quad P = \frac{\sum_{t}TP_{t}}{\sum_{t}(TP_{t}+FP_{t})},R = \frac{\sum_{t}TP_{t}}{\sum_{t}(TP_{t}+FN_{t})}\\
&\quad \quad \quad \quad {\rm Micro-F1} = \frac{2PR}{P+R}\\
\end{aligned}
\end{equation}

Both Macro-F1 and Micro-F1 are informative metrics. The former gives the performance on each category an equal weight in computing the average; the latter gives the performance on each instance an equal weight in computing the average. In fact, Micro-F1 is the  accuracy.

\subsection{Evaluation of CPTDL}
To demonstrate the effectiveness of CPTDL, we compare our method against the following baselines based on two kinds of VGG16 Networks, namely ImageNet CNN and Places CNN:
\begin{myitemize}
\item Video-O: Directly using  key frames from  training videos to fine-tune the ImageNet CNN.
\item Image-O: Directly using all the Foursquare images to fine-tune the ImageNet CNN.
\item Video-Image-O: first using key frames from  training videos, and then using  Foursquare images  to fine-tune the ImageNet CNN.
\item Image-Sel-O: Using the selected Foursquare images to fine-tune the ImageNet CNN.
\item O-Late Fusion: Using the selected Foursquare images and key frames from  training videos  separately to fine-tune two ImageNet CNNs, and  then using the max pooling  scores as the final prediction.
\item Video-S: Directly using key frames from  training videos to fine-tune the Places CNN.
\item Image-S: Directly using Foursquare images to fine-tune the Places CNN.
\item Video-Image-S: First use key frames from  training videos, and then using  Foursquare images  to fine-tune the Places CNN.
\item Image-Sel-S: Directly using the selected Foursquare images to fine-tune the Places CNN.
\item S-Late Fusion: Using the selected Foursquare images and videos  separately to fine-tune two Places CNNs, and then using the max pooling  scores as the final prediction.
\end{myitemize}

Our method CPTDL-O and CPTDL-S first use  key frames from  training videos to fine-tuned the network, and then  use the fine-tuned model to select foursquare images for further fine-tuning.

The comparative results are summarized in Table \ref{cntdl_results}. From Table \ref{cntdl_results}, we can observe four key findings: (1) Performance can be improved by
taking advantage of Foursquare images for venue category prediction from videos in both two types of networks. Particularly, for ImageNet CNN, compared with Video-O, CPTDL-O achieves a performance improvement of about 1.2 percent in Macro-F1 and 4 percent in Micro-F1, respectively. For Places CNN, there is also the performance improvement of about 0.2 percent in Macro-F1 and 1.1 percent in Micro-F1, respectively. This validates the effectiveness of using images from other platforms to boost the prediction performance of videos in Vine. (2) Our proposed CPTDL performs better than other baselines (e.g., Video-Image-O, O-Late Fusion) that use both images and videos, which validate that our method is effective in learning more discriminative features by taking full advantage of  images from Foursquare. (3) We can see that the performance of CPTDL-S is better than CPTDL-O, which indicates that  the high-level scene  information is more discriminative  than the object  information for the task of venue category prediction. This is reasonable, since the venue is location-sensitive and  scene information is more important than  object information. (4) Video-Image-O and Video-Image-S do not perform better than Video-O and Video-S. This is because images from Foursquare are very noisy and  may have the semantic drift for a video venue category, which will lead the fine-tuning to the wrong direction. Note that Image-Sel-O and  Image-Sel-S perform worse than other methods. The reason is that after image filtering, there are no images for some venue categories and thus the accuracy on these venue categories is 0.
\begin{table} [t]
\caption{Performance comparison between our method and the baselines in CPTDL}
\label{cntdl_results}       
\centering
  \begin{tabular}{|l|l|l|l|}
  \hline
    \textbf{Method}& Macro-F1 & Micro-F1\\
    \hline
    Video-O&13.90\%&28.52\% \\
    \hline
    Image-O&12.30\%&16.50\% \\
    \hline
    Video-Image-O&13.97\%&28.51\%\\
    \hline
    Image-Sel-O&8.30\%&12.50\%\\
    \hline
    O-Late Fusion&13.98\%&28.72\%\\
    \hline
    CPTDL-O&\textbf{15.08\%}&\textbf{32.50\%}\\
    \hline
    \hline
    Video-S&15.27\%&30.07\%\\
    \hline
    Image-S&13.50\%&16.70\%\\
    \hline
    Video-Image-S&15.24\%&30.10\%\\
    \hline
    Image-Sel-S&8.50\%&12.90\%\\
    \hline
    S-Late Fusion&15.30\%&31.05\%\\
    \hline
    CPTDL-S&\textbf{15.50\%}& \textbf{33.60\%}\\
    \hline
  \end{tabular}
\end{table}

\subsection{Evaluation of Multi-View Feature Learning (MVFL)}
After CPTDL, we obtain object and scene features from CPTDL-O and CPTDL-S. Based on the extracted object-scene features, we further verify the effectiveness of MVFL without the venue hierarchical structure prior.
We consider the following baselines for comparison:
\begin{myitemize}
\item Object-Feature (O-Fea). This baseline first extracts the visual features of key frames using the  fine-tuned ImageNet CNN from CPTDL, then conduct a mean pooling to obtain the final video representation.
\item Scene-Feature (S-Fea). Similar to O-Fea, but use the  fine-tuned-Places CNN from CPTDL.
\end{myitemize}

\begin{table} [t]
\caption{Performance comparison between our method and the baselines in MVFL}
\label{cnmvdsf_results}       
\centering
  \begin{tabular}{|l|l|l|l|}
  \hline
    \textbf{Method}& Macro-F1 & Micro-F1\\
    \hline
    O-Fea& 15.08\% &32.50\% \\
    S-Fea&15.50\%&33.60\%\\
  \hline
    MVFL-1-4096&15.88\%&33.40\%\\
    MVFL-2-4096&15.92\%&34.20\%\\
    MVFL-3-4096&15.89\%&33.80\%\\
    MVFL-1-8192&16.92\%&34.60\%\\
    MVFL-2-8192&\textbf{17.28\%}&\textbf{35.20\%}\\
    MVFL-3-8192&16.93\%&34.70\%\\
    \hline
  \end{tabular}
\end{table}
MVFL network fused object features and  scene features into a joint feature representation via a multi-layer feed-forward  network. We use  MVFL-L-D to denote the MVFL network with $L$ fused layers and the units of each fused layer is $D$. For example, MVFL-1-4096 denotes there is a  joint layer with the 4,096 units.

The comparative results are summarized in Table \ref{cnmvdsf_results}. We can see that these two kinds of features are complementary. Combining them offers better performance than single features in the task of venue category predication. We further observe that when the number of joint layers is 2 and the units of each joint layer is 8,192, the performance is the best. Therefore, we will use such deep architecture to  evaluate our framework.

\subsection{Evaluation of the HCM-FL}
Combining CPTDL and MVFL-HVS, we finally verify the effectiveness of our proposed HCM-FL framework. Considering the task of venue category prediction, we choose the following methods as comparison:
\begin{myitemize}
\item TRUMANN \cite{Zhang-VCEMV-MM2016}: This is the first work to introduce the task of venue category predication from videos. This method proposed a tree-guided multi-task multi-modal learning model for venue category prediction.
\item DARE \cite{Nie-EMVU-MM2017}: This baseline  compensated  the acoustic modality via harnessing external sound knowledge and developed a deep transfer model for venue category prediction.
\end{myitemize}
\begin{table} [t]
\caption{Performance comparison between our HCM-FL and other methods on venue category prediction}
\label{CNMVTDL_results}       
\centering
  \begin{tabular}{|l|l|l|l|}
  \hline
    \textbf{Method}& Macro-F1 & Micro-F1\\
    \hline
    TRUMANN \cite{Zhang-VCEMV-MM2016}& 5.21\% & 25.27\%\\
    DARE \cite{Nie-EMVU-MM2017}& 16.66\% & 31.21\%\\
    \hline
    MVFL-2-8192&{17.28\%}&{35.20\%}\\
    HCM-FL&\textbf{18.82\%}&\textbf{37.40\%}\\
    \hline
  \end{tabular}
\end{table}

Table \ref{CNMVTDL_results} shows the experimental results. We can see that after introducing the hierarchical structure prior, the performance has been  further improved than MVFL-2-8192. Furthermore, compared with DARE, our HCM-FL has a significant performance gain. HCM-FL can improve the relative performance by 13\% for Macro-F1 and 20\% for Micro-F1, respectively. This verifies the effectiveness of HCM-FL in jointly utilizing the multi-platform data, multi-view deep features and the hierarchical venue structure prior.

\begin{figure}
\centering
\includegraphics[width=0.40\textwidth]{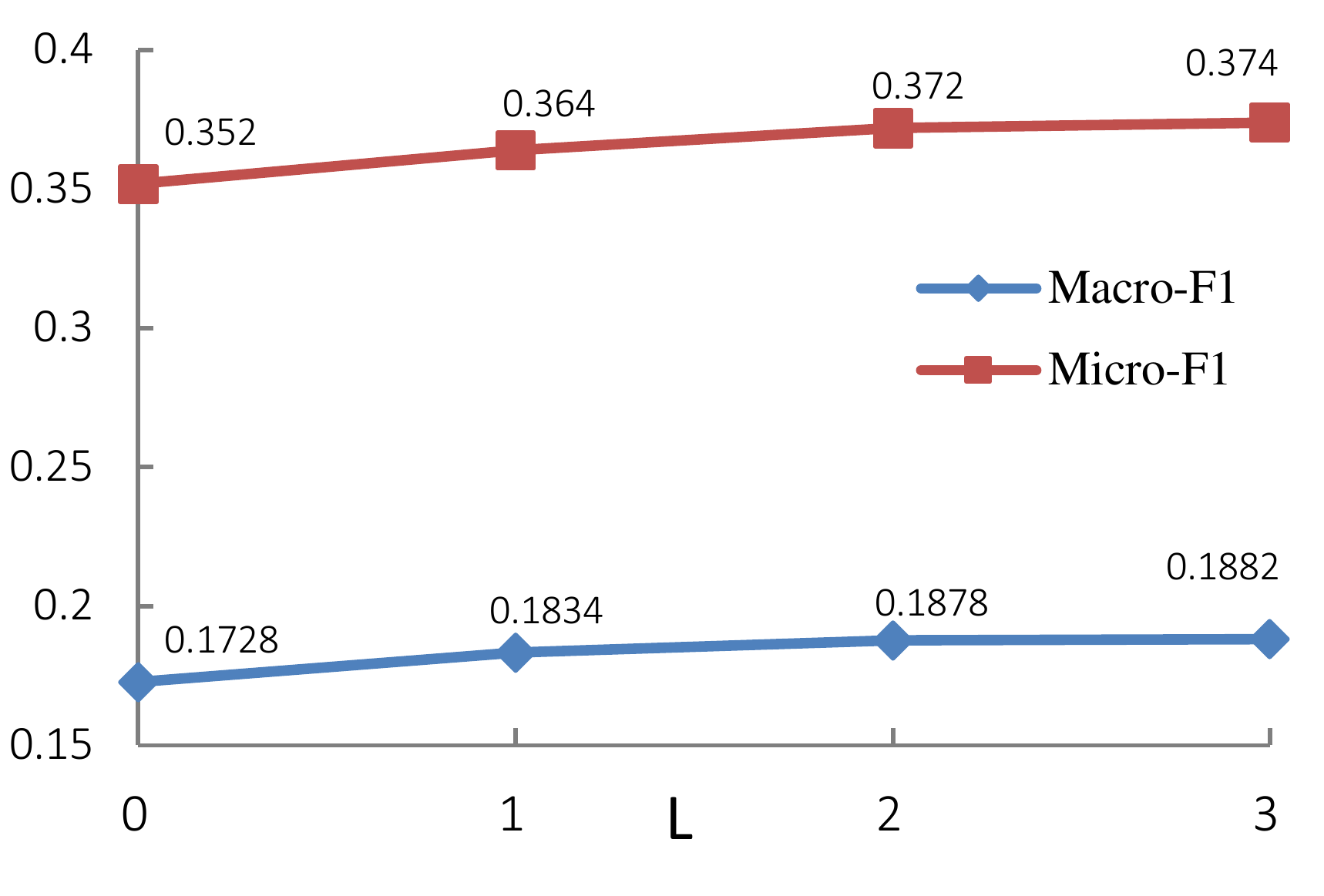}
\caption{Comparison of our methods with different hierarchical venue layers.}
\label{performance_variance_L}
\end{figure}

Figure \ref{performance_variance_L} further compares our framework with different number $L$ of hierarchical layers, where $L=\{0,1,2,3\}$. We can see that  there is a consistent increase for  both Macro-F1 and Micro-F1  as we increase the layers of venues $L$ from 0 to 3. This shows that representations with higher venue layers become increasingly better at discovering useful features.  The reason is that introducing more hierarchical layers makes HCM-FL utilize more prior knowledge to improve the performance. We can also see that the increasing amplitude of the performance becomes smaller with the increase of venue layers. The probable reason is  higher venue layers give more minimal effect in improving the prediction performance.

\begin{figure*}[!t]
\centering
\subfigure[]{\includegraphics[width=0.90\textwidth]{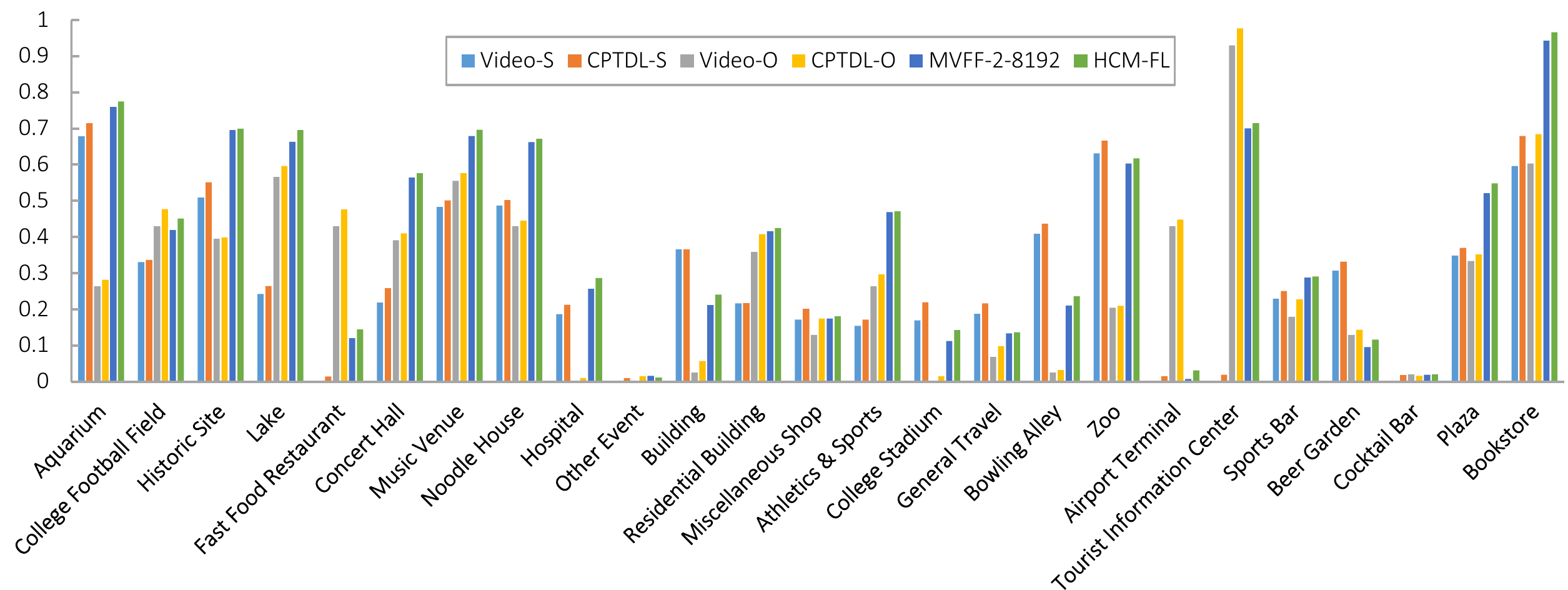}%
}
\vfil
\subfigure[]{\includegraphics[width=0.90\textwidth]{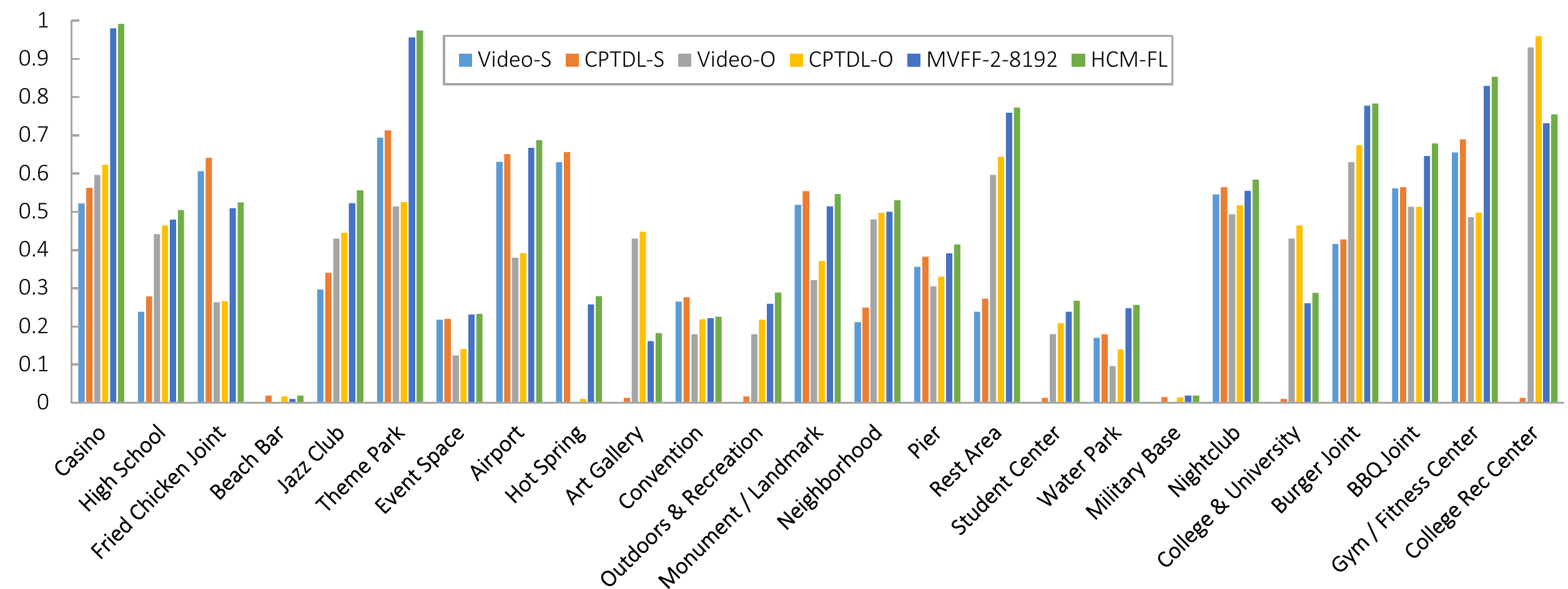}%
}
\caption{Per-venue category prediction results for HCM-FL  on 50 venue categories}
\label{singe_venue_category}
\end{figure*}

For additional analysis, we also provide venue category-specific results from HCM-FL in Fig. \ref{singe_venue_category}. For the space limit, we show the results on 50 venue categories from 188 ones. We report the number that using  the following six methods: Video-S, CPTDL-S, Video-O, CPTDL-O, MVFL-2-8192 and HCM-FL. We observe that (1) After CPTDL, there is a consistent performance gain than without transfer deep learning for all the 50 venue categories.  The performance of CPTDL-S is better than Video-S, similarly for Video-O and CPTDL-O. This further confirmed that our proposed framework can better leverage the  strength of images from other platforms to improve the performance.  (2) The scene and object features are complementary. Therefore, the performance of  MVFL-2-8192 is better than CPTDL-S and CPTDL-O  for many venue categories. There are some failure cases, such as the venue category of Fast Food Restaurant. The reason is that there is a relative large difference for the performance of CPTDL-S and CPTDL-O. Therefore, their fusion leads to an intermediate trade-off between these two features in the prediction performance. Take the venue category Fast Food Restaurant as an example, the performance of CPTDL-O is 1.48\% while the performance of CPTDL-S is 47.62\%. The performance of MVFL-2-8192 is 12.10\%. (3) After introducing the hierarchical venue prior, there is a performance gain for  most venue categories out of 50 venue categories.

\subsection{Discussions}
\begin{figure}
\centering
\includegraphics[width=0.35\textwidth]{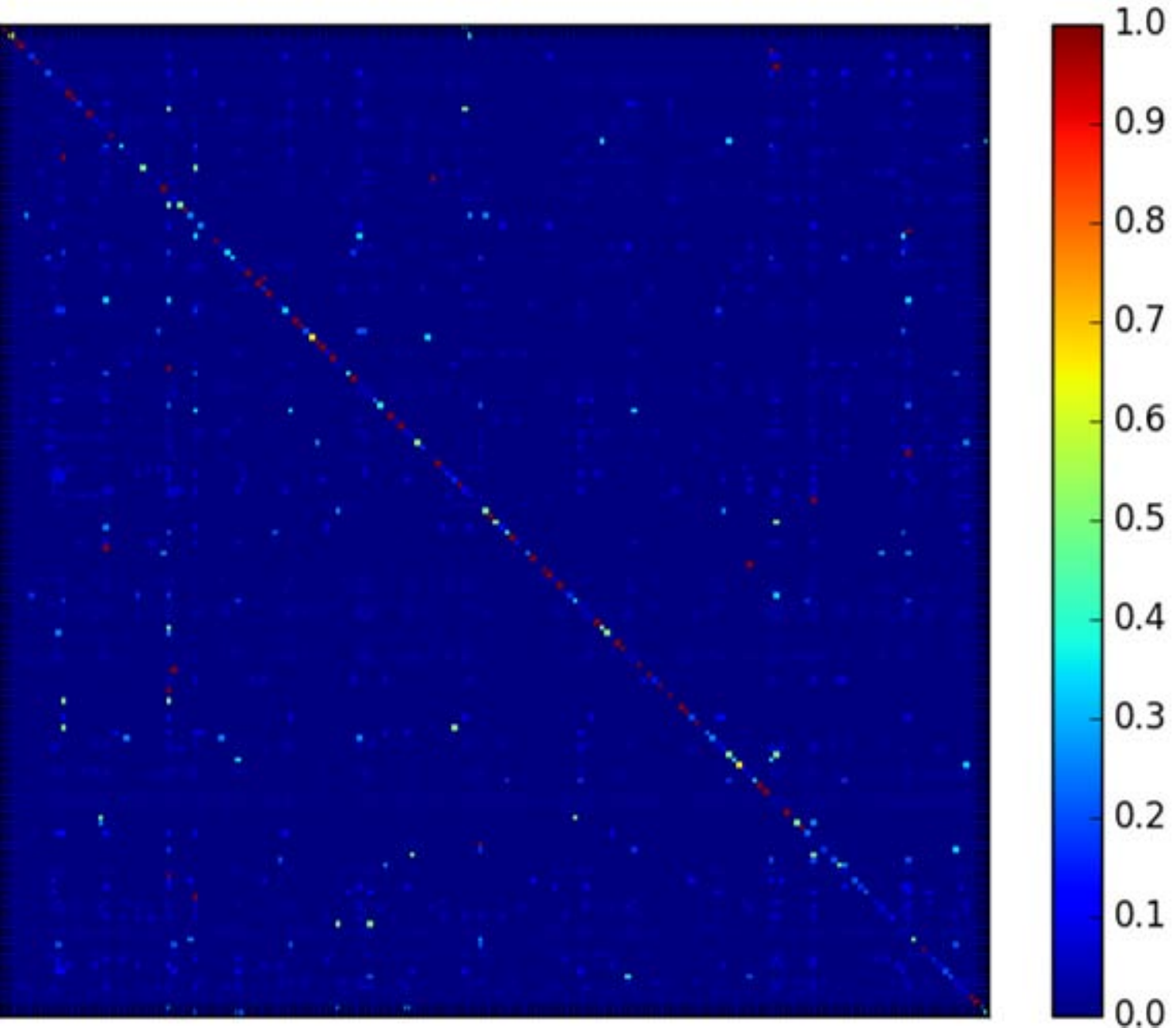}
\caption{The detailed comparison  over each individual venue category for HCM-FL via the confusion matrix. The row denote true label and the column denote the estimated label. (best viewed under magnification).}
\label{confusion_matrix}
\end{figure}

\begin{figure*}
\centering
\includegraphics[width=0.90\textwidth]{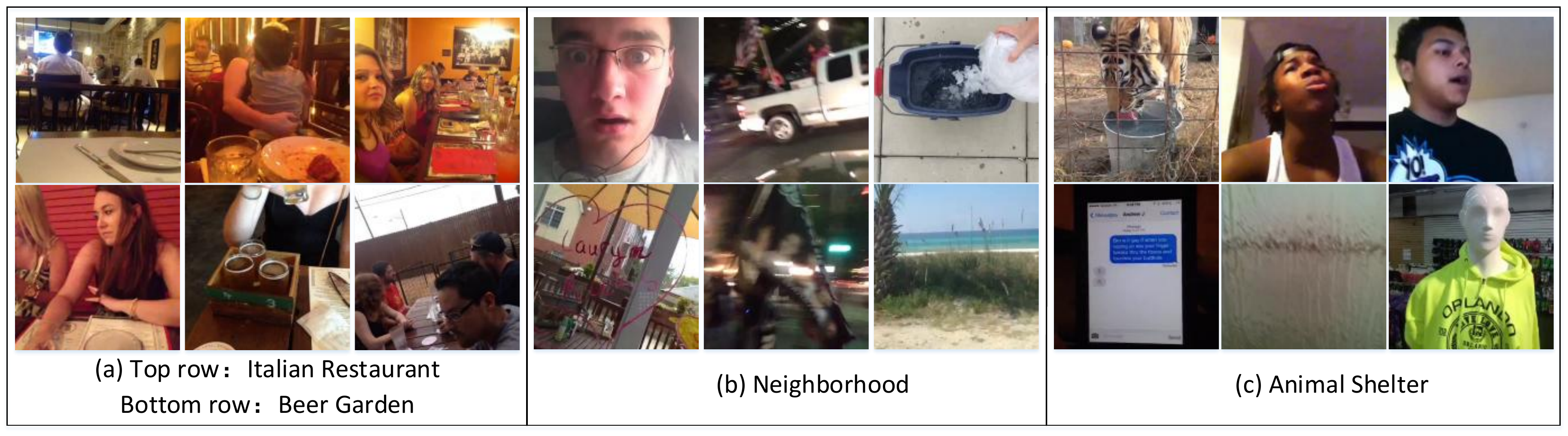}
\caption{Some videos from different venue categories to show (a) low inter-class variation, (b) high intra-class variation and (c) some wrong-labeled venue categories}
\label{discussion_examples}
\end{figure*}

Our proposed framework has the relative higher performance gain than existing methods for venue category prediction from videos in Vine. However, there are  still relative low prediction performance. This section lays out  additional observations that follow from our results to find  the probable reasons.

Fig. \ref{confusion_matrix} showed  the confusion matrix of  HCM-FL over each individual venue category. We can see that our method still does not provide perfect accuracy for some  venue categories. We further observe the video data with lower prediction performance and  find the following several reasons:  (1) Low inter-class variation. For example, as shown in Fig. \ref{discussion_examples} (a), the videos to describe the Italian Restaurant and  Beer Garden  venue category are visually similar.  Therefore,  many test videos from the Italian Restaurant  are misclassified into the Beer Garden venue category. In our experiment, the ratio of videos, which are classified to the Beer Garden  is about 33.33\% while the accuracy for the Italian Restaurant is near 0. (2) High intra-class variation. For example, as shown in Fig. \ref{discussion_examples} (b),  the intra-class variation  for the neighborhood venue category is too large. (3) Wrongly-labeled videos. There are also some wrong-labeled videos. Fig. \ref{discussion_examples} (c) shows some wrongly-labeled videos from the venue category ``Animal Shelter". (4) Too few training samples. For example, the number of videos for the Beach Bar is totally  only 61. Therefore, its accuracy  on Video-O and Video-S is 0. Although after transfer deep learning from Foursquare images, there is an improvement. However, the performance is still low, namely 1.71\% and 1.90\%, respectively. In order to relieve the above problems, it is probably helpful to  combine  visual information with  textual and other modality information into our framework to solve problem (1) and (2). For  problem (3), there are some existing solutions such as \cite{Xiao-LMNLD-CVPR2015,Vahdat-RLN-arXiv2017} to refer and they  proposed a  solution to train CNNs when there exist mislabeled images in the training set. For problem (4), we can also resort to visual information from more platforms to add more training data.
\section{Related Work}
Our work is closely related to the following three research areas: (1) location recognition and prediction, (2) cross-platform analysis and applications, and (3) feature learning with the hierarchial class structure.

\subsection{Location Recognition and Prediction}
The goal of location recognition and prediction is to assign the location information to the given  text, image or video. It can support a variety of  applications, such as  event discovery \cite{Dredze-GLT-NAACL2016}, advertising, personalization, location recommendation \cite{Farseev-CDR-SIGIR2017} and location visualization \cite{Sang-ESMI-TIST2017}. According to the level of location granularity, there are mainly four types of location prediction, including GPS-level, POI-level, city-level and country-level location  prediction.

The  task of GPS-level location prediction is to estimate the GPS location  given the text \cite{Hulden-KDE-AAAI2015}, images \cite{Hays-IM2GPS-CVPR2008,Vo-RIM2GPS-ICCV2017}  or multi-modal information \cite{Trevisiol-RGLV-ICMR2013,Choi-MMLE-Springer2015}. For example, Song \textit{et al.} \cite{Song-WVG-TMM2012} propagated geotags among the web video social relationship graph for video geo-location. In contrast, in many real-world scenarios, especially in  social media applications, it is more important for POI-level location prediction. For example, Chen \textit{et al.} \cite{Chen-MBVR-AAAI2016} mined business-aware visual concepts from social media for recognizing the business venue  of  images. In addition, there are many retrieval based methods for location estimation \cite{Hays-IM2GPS-CVPR2008,Shuang-INSTRE-TOMCCAP2015,Weyand2016PlaNet,Vo-RIM2GPS-ICCV2017}. The first work on venue category prediction from videos is from Zhang \textit{et al.} \cite{Zhang-VCEMV-MM2016}. They proposed a tree-guided multi-task multi-modal learning model for venue category prediction. In addition, they released a  large-scale micro-video dataset. Recently, Nie \textit{et al.}  \cite{Nie-EMVU-MM2017}   fused multi-modal information, especially the acoustic modality for venue category prediction from videos.  Among POI-level location prediction, visual landmark recognition and analysis \cite{Zheng-TWBWSLR-CVPR2009,Min-MLS-TMM2014,Min-MSTTM-IEEEMM2014,Guan-MLR-TIST2015} has also been widely studied for its tourism applications. There are also some works  on  city-level \cite{Han-TTUGP-JAIR2014} or country-level location prediction \cite{Zubiaga-TRTCL-TKDE2016}. Similar to \cite{Zhang-VCEMV-MM2016,Nie-EMVU-MM2017}, our work belongs to the POI-level venue category prediction. However,  we  focus on cross-platform venue category prediction from videos, where the media data from other platforms are effectively exploited to improve the prediction performance.

\subsection{Cross-Platform  Analysis and Applications}
With the fast development of Web2.0, various  media-sharing platforms are gaining more and more popularity with their different types of data and services. Therefore, more and more works resort to cross-platform based study for various applications, such as  cross-network based recommendation \cite{Min-CPMMTM-TMM2015,Yan-CNA-TMM2015}, event detection \cite{Bao-CPETD-TOMM2015}, popularity prediction \cite{Roy-CDLSVPP-TMM2013}, video recognition and retrieval \cite{Gan-YLWE-CVPR2016,Han-VRFP-TMM2016}. For example,  Roy \textit{et al.} \cite{Roy-CDLSVPP-TMM2013} proposed a novel transfer learning framework that utilizes the knowledge from Twitter to grasp sudden popularity bursts in online content from Youtube. Min \textit{et al.} \cite{Min-CPMMTM-TMM2015} conducted the recommendation between  two platforms, namely  photo recommendation from Flickr to Foursquare users and venue recommendation from Foursquare to Flickr users. Gan \textit{et al.} \cite{Gan-YLWE-CVPR2016} presented a  labor-free video concept learning framework by jointly utilizing noisy web videos from Youtbue and images from Google. Different from them, we take advantage of the images from Foursquare for venue category prediction from videos in Vine. In addition, we also jointly utilized multiple semantic cues (e.g., scenes and objects) and the venue structure  prior to boost the prediction performance.

\subsection{Feature Learning with the Hierarchical Class Structure}
More recently, driven by the great success of Convolutional Neural Networks (CNN) on image recognization tasks \cite{Alex-ImageNet-NIPS2012,Simonyan-VDCN-arXiv2014}, a few works attempted to leverage  CNN models to learn  feature representations for location recognition. For example,  Weyand \textit{et al.} \cite{Weyand-PlaNet-arXiv2016} proposed  a deep image classification approach in which the world is spatially divided into cells and a deep network is trained to predict the correct cell for a given image.  All these works, however, focus on extracting  visual or multi-modal features  using neural networks, and do not utilize the external class structure knowledge.

Utilizing  the hierarchical class structure has become increasingly important for its ability to learn more discriminative features, especially under unbalanced sample distributions over different classes \cite{Srivastava-DTLTP-NIPS2013}. For example, Gopal \textit{et al.} \cite{Gopal-LSHC-NIPS2012} proposed a  Bayesian method to model hierarchical dependencies among class labels using  multivariate logistic regression. Different from \cite{Gopal-LSHC-NIPS2012}, some works \cite{Srivastava-DTLTP-NIPS2013,Fu-TPPTDL-WWW2015}  combined the strength of  deep neural networks, with tree-based priors, making the deep neural networks work well on unbalanced class distributions. Wang \textit{et al.} \cite{Wang-VSCSIT-AAAI2016} exploited  hierarchically structured tags from  different abstractness of semantics and multiple tag statistical correlations, thus discovered more accurate semantic correlations among different video data, even with highly sparse and incomplete tags. There is a natural geographically hierarchical structure  for location information. Some works such as \cite{Zhang-LGHF-IJCAI2015,Zhang-LGHF-TMM2016} have explored such prior  for  image location prediction. Zhang \textit{et al.} \cite{Weyand-PlaNet-arXiv2016} proposed a  tree-guided multi-task multi-modal learning approach  to jointly fuse multimodal information, including  deep visual features, textual features and audio features  from videos for venue category prediction. Different from their work, we use the Bayesian model to exploit the pre-defined hierarchical venue structure prior and combine it with multi-view feature fusion network and  transfer deep learning from other platforms to learn more discriminative deep features  for venue category prediction.

\section{Conclusions}

We have proposed a Hierarchy-dependent Cross-platform Multi-view Feature Learning (HCM-FL) framework, which  jointly utilized  multi-platform data, object-scene deep features and
the hierarchical venue structure prior for venue category prediction from videos. In order to utilize the multi-platform data effectively, we proposed a  cross-platform transfer deep learning method, which leveraged the complementary nature of the media data from two platforms to reinforce the learned deep network from videos in Vine using the images from Foursquare. In addition, HCM-FL  augmented the object-scene feature fusing  network with the hierarchical venue structure prior to enable the fusing  network to transfer knowledge from related venue categories. Experiments show that HCM-FL achieves better results on  two datasets  from Vine and Foursqure  than existing methods.

As discussed earlier, our work can be extended in the following three directions: (1) We have found that some videos in Vine are wrongly labeled. Therefore, how to improve the robustness of our model under noisy labels is our first direction. Some works \cite{Xiao-LMNLD-CVPR2015,Vahdat-RLN-arXiv2017} proposed a  solution to train CNNs when there exist mislabeled images in the training set. (2) We can extend our framework to use  data from more platforms.  (3) Multi-modal information (e.g.,  text and audio information) from multiple platforms can  also be exploited in the future. In addition, besides the appearance information, temporal information from videos also contains discriminative signals. Thus, we also plan to use the  LSTM to  capture the sequential features for venue category prediction \cite{Meng_MVU_LiuNWC17}.

\bibliographystyle{abbrv}
\bibliography{TMM2018ref}  

\end{document}